\documentclass[journal,twocolumn]{IEEEtran}
\usepackage{}
\usepackage{graphicx}
\usepackage{epstopdf}
\usepackage{amssymb}
\usepackage{amsfonts}
\usepackage{amsmath}
\usepackage{algorithm}
\usepackage{algorithmic}
\usepackage{subeqnarray}
\usepackage{cases}
\usepackage{color}
\usepackage{amsmath,lipsum}
\usepackage{cuted}

\usepackage{bm}
\usepackage{subfigure,amsmath,amssymb,cite}

\usepackage{float}
\ifCLASSINFOpdf
\else
\fi
\begin{document}

\title{A New Heterogeneous Hybrid Massive MIMO Receiver with An Intrinsic Ability of Removing Phase Ambiguity of DOA Estimation via Machine Learning}
%
\author{ 
Feng Shu, Baihua Shi, Yiwen Chen, Jiatong Bai, Yifan Li, Tingting Liu, \\Zhu Han, \IEEEmembership{IEEE, Fellow}, and Xiaohu You, \IEEEmembership{IEEE, Fellow}
\thanks{This work was supported in part by the National Natural Science Foundation of China (Nos.U22A2002, 62171217 and 62071234), the Hainan Province Science and Technology Special Fund (ZDKJ2021022), the Scientific Research Fund Project of Hainan University under Grant KYQD(ZR)-21008, and the Collaborative Innovation Center of Information Technology, Hainan University (XTCX2022XXC07). (Corresponding authors: Feng Shu).}
\thanks{Feng Shu is with the School of Information and Communication Engineering and Collaborative Innovation Center of Information Technology, Hainan University, Haikou 570228, China, and also with the School of Electronic and Optical Engineering, Nanjing University of Science and Technology, Nanjing 210094, China (e-mail: shufeng0101@163.com).}
\thanks{Baihua Shi and Yifan Li are with the School of Electronic and Optical Engineering, Nanjing University of Science and Technology, Nanjing 210094, China.}
\thanks{Yiwen Chen and Jiatong Bai are with the School of Information and Communication Engineering, Hainan University, Haikou, 570228, China.}
\thanks{Tingting Liu is with the School of Computer Science, Nanjing University of Posts and Telecommunications, Nanjing 210003, China.}
\thanks{Zhu Han is with University of Houston, Houston, TX 77004, USA.}
\thanks{Xiaohu You is with the National Mobile Communications Research Laboratory, School of Information Science and Engineering, Southeast University, Nanjing 210096, China, and also with the Purple Mountain Laboratories, Nanjing 211111, China.}

}
\maketitle
\begin{abstract}
	Massive multiple input multiple output (MIMO) antenna arrays eventuate a huge amount of circuit costs and computational complexity. To satisfy  the needs of high precision and low cost in future green wireless communication, the conventional Hybrid analog and digital MIMO receive structure emerges a natural choice. But it exists an issue of the phase ambiguity in direction of arrival (DOA) estimation and requires at least two time-slots to  complete one-time DOA measurement with the first time-slot generating the set of candidate solutions and the second one to find a true direction by received beamforming over this set, which will lead to a low time-efficiency. To address this problem, 
	a new heterogeneous sub-connected  hybrid analog and digital ($\mathrm{H}^2$AD) MIMO  structure is proposed with an intrinsic ability of removing phase ambiguity, and then a corresponding new framework is developed to implement a rapid high-precision DOA estimation using only single time-slot. The proposed framework consists of two steps: 1) form a set of candidate solutions using existing methods like MUSIC; 2) find the class of the true solutions and compute the class mean. To infer the set of true solutions, we propose two new clustering methods: weight global minimum distance (WGMD) and weight local minimum  distance (WLMD). Next, we also enhance two classic clustering methods: accelerating local weighted k-means (ALW-K-means) and improved density.
	Additionally, the corresponding closed-form expression of Cramer–Rao lower bound (CRLB) is derived.
	Simulation results show that the  proposed frameworks using the above four clustering can approach the CRLB at almost all signal to noise ratio (SNR) regions except for extremely low SNR (SNR$<-5$ dB). Four clustering methods have an accuracy decreasing order as follows: WGMD, improved DBSCAN, ALW-K-means and WLMD.
\end{abstract}


\section{Introduction}
Target localization is a well attended issue in wireless communication research and is widely used in many fields, which mainly includes the techniques of time of arrival (TOA), received signal strength (RSS) and angle of arrival (AOA)\cite{Handbook-19,shiAOA2023drones,wang2015twc,wang2018twc}. In particular, the authors discussed a methodology to achieve optimal geometric configurations under regional constraints for the RSS-based localization problem of Unmanned Aerial Vehicles (UAVs) in \cite{9917309}. Direction of Arrival (DOA) technique is a key part of target localization, which can provide precise desired signal direction for many advanced wireless communication technologies, like directional modulation \cite{shuSPWT,Shunetwork}, reconfigurable intelligent surfaces (RIS) aided communication \cite{WuIRS,tengTcom2021RIS,wangRIS2023ojcs}, et. al. What's more, integrated sensing and communications (ISAC)  is in vigorous development and has been a key part of the sixth generation (6G). DOA estimation is also a indispensable part in ISAC, which can be used in vehicle to everything, smart home, remote sensing, and so on \cite{cuiISAC2021network}.

There are four main kinds of methods in DOA estimation. The first is traditional estimation methods, such as beamforming, maximum likelihood estimation (MLE), and fast Fourier transform (FFT). These methods cannot estimate targets in close proximity. Therefore, algorithms of the second kind, subspace based methods, are more popular. These methods are able to distinguish adjacent angles by using the orthogonality of signal subspace and noise subspace, such as multiple signal classification(MUSIC) \cite{MUSIC} and Estimation of signal parameters via rotational invariance technique (ESPRIT) \cite{ESPRIT}. Based on the two algorithms above, many modified high performance and low complexity methods were developed \cite{ESPRIT,rootmusic,urootMUSIC,TLSESPRIT2}.
The third is compressed sensing (CS) based methods \cite{MalioutovCS1}. Due to the sparse of DOA, CS is able to estimate the angles. Furthermore, CS can reconstruct received signals to deal with one-bit ADCs and nonuniform arrays \cite{wangCSDOA,zhouCSDOA}. Especially, atomic norm has been adopted to increase the degrees of freedom in co-prime arrays \cite{zhengCoprime,zhangCoprime}. The last approach is based on machine learning. Due to the high computational complexity of traditional algorithms, especially in massive Multiple Input Multiple Output (MIMO), machine learning is employed to decrease that \cite{heDOA2018wcl,huangDOA2018tvt,maDOA2024spl}. In \cite{huangDOA2018tvt}, a deep neural network (DNN) was used to conduct DOA estimation. In addition, multiple DNN and convolutional Neural Networks (CNN) were adopted in \cite{huDOA20020wcl} and \cite{maDOA2024spl}, respectively. 

In recent years, the emergence of MIMO technology in combination with DOA can achieve ultra-high angular resolution, but on the contrary, it brings a significantly higher RF-chain circuit cost and computational complexity. 
To solve this challenge, adopting low-resolution analog to digital convertors (ADCs) is a promising solution. Low-resolution ADCs can save considerable circuit cost and power consumption. There are many existed works about low-resolution ADCs, like one-bit ADC, low-resolution ADC structure, mixed-ADC structure, and so on \cite{Zhangjsac,onebitMUSIC,Zheng2022onebit,mixedonebit,SBH-ADC,shiDOA2023ojcs,heDOA2018jstsp}. 
The uplink achievable rate of mixed-ADC structure was analyzed in \cite{Zhangjsac}.
In \cite{onebitMUSIC}, authors proved that the multiple signal classification (MUSIC) method can be used in one-bit ADC structure without modification. In \cite{mixedonebit}, atomic norm was adopted to measure the DOA in mixed one-bit antenna array. Afterwards, the performance and energy efficiency were investigated in \cite{SBH-ADC}. Authors also proved that all subspace-based methods can be used in mixed-ADC architecture.

Hybrid analog and digital (HAD) structure connect one radio frequency (RF) chain with multi antennas. Every antenna has a analog phase shifter. Then the signals from multi antennas are added and converted to the baseband signals by one RF chain. Thus, There are fewer RF chains and ADCs in HAD structure, which can save much circuit cost and energy consumption. 
Many researches have been donated to the DOA estimation in HAD structure \cite{huDOA20020wcl,liDOA2020wcl,shuHADDOA2018tcom,fanDOA2018twc,zhangDOA2022tsp}.
In \cite{shuHADDOA2018tcom}, four low-complexity methods for sub-connect HAD array were proposed and the corresponding Cramer–Rao lower bound (CRLB) was derived as a benchmark. A 2-D fully-connect HAD was considered in \cite{fanDOA2018twc}, and a 2-D discrete Fourier transform based method was proposed to estimate the DOA. 
A dynamic maximum likelihood (D-ML) estimator was derived for the DOA estimation of fully-connected, sub-connected, and switches-based HAD structure in \cite{zhangDOA2022tsp}
In \cite{shiDOA2023ojcs}, the sub-connect HAD structure with mixed-ADC was investigated. Authors proposed a novel method and proved that the proposed structure have higher energy efficiency.

\begin{figure*}[!htb]
	\centering
	\subfigure[Proposed $\mathrm{H}^2$AD array with $Q$ groups, where group $q$ has $K_q$ subarrays with each subarray having $M_q$ antennas, $q\in S_Q=\{1,2,\cdots,Q\}$ ]{\includegraphics[height=1.65in]{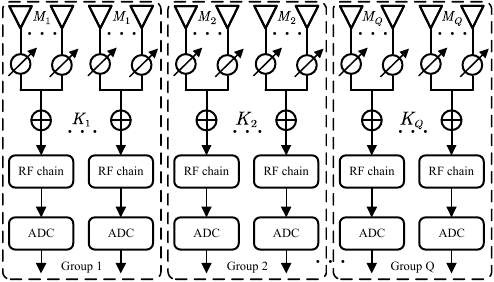}\label{fig: sub_figure1}}
	\hspace{0.4cm}
	\subfigure[Conventional sub-connect HAD array ]{\includegraphics[height=1.65in]{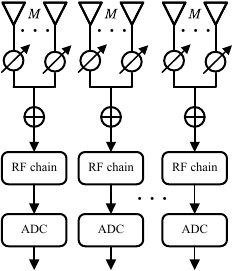}\label{fig: sub_figure2}}
	\hspace{0.4cm}
	\subfigure[Conventional fully-digital array]{\includegraphics[height=1.65in]{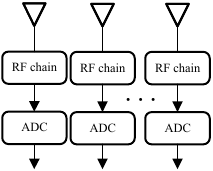}\label{fig: sub_figure3}}
	\caption{Three different antenna arrays}
	\label{fig_sys_mod}
\end{figure*}

Conventional fully-digital massive or extremely massive MIMO with uniform linear array (ULA) may achieve an extremely high- performance DOA measurement. However, its main shortcoming are high circuit cost and computational complexity. Consequently, the HAD structure emerged. In \cite{shuHADDOA2018tcom}, the authors proved that the HAD can achieve a high performance with a low complexity and circuit cost. However, sub-connected HAD structure is subjected to phase ambiguity, which is caused by interval of multiple wavelengths between adjacent subarrays. As a result, it requires at least several sample time-slot to complete one-time DOA measurement. This will result in  a low-time efficiency and is particularly unsuitable for the future 6G latency-sensitive applications. In general, during the first time-slot, a receive sample signal sequence is adopted to generate a set of candidate angles. Then,  one or more  new receive signal sequences  with receiving beamforming alignment or scanning over this set are sampled to remove the false solutions and find the the true solution. The low time-efficiency will limit its future wide applications.

To address the problem of low time efficiency, in this paper, a new heterogeneous sub-connected  hybrid analog and digital ($\mathrm{H}^2$AD) massive MIMO receiver structure is proposed with an intrinsic ability of removing phase ambiguity of DOA estimation.
The main contributions of our work can be summarized as follows:
\begin{enumerate}
	\item To achieve a high time-efficiency, a new $\mathrm{H}^2$AD receiver structure is proposed, and  is composed by $Q$ subarray groups, where each subarray group is a conventional sub-connect HAD structure with each subarray including the same number of antennas. For different groups, the associated subarrays are employed with different numbers of antennas. To make a rapid DOA estimate, based on this structure,  a two-steps framework is proposed as follows: 1) generate the set of candidate solutions by adopting conventional DOA estimation methods, like root-MUSIC, etc. Here, the set is composed of two clusters: false solutions and true solutions. 2) infer the cluster of true solutions and merge the true solutions. The set of candidate solutions play a bridge role between traditional statistical signal processing (Step 1) and advanced machine learning (Step 2). In the following, we will focus on how to design high-performance classification methods to find the class of true solutions.
	\item  To make a high-accuracy classification of the set of candidate solutions, a high-performance clustering method, called weight global minimum distance clustering (WGMD) is proposed. To reduce the high computational complexity of WGMD, a weight local minimum distance clustering (WLMD) is designed and two conventional clustering methods are enhanced as follows: accelerated k-means method and improved density-based spatial clustering of applications with noise (DBSCAN). 
	Finally, a minimum mean squared error (MMSE) based method is derived to merge all elements of the class of true solutions by taking the different DOA measurement precision of each group due to the fact that the number of subarrays of each group varies over groups. 
	Simulation results showed that all methods could achieve a $100\%$ accuracy at almost all SNR regions except for extremely low SNR region ($<-5$ dB), an accuracy decreasing order is listed as follows: WGMD, improved DBSCAN, ALW-K-means and WLMD.
	\item To evaluate the performance of the proposed structure, framework, and methods,  the closed-form expression of the corresponding CRLB for the $\mathrm{H}^2$AD structure is derive and  may be regarded as a benchmark. In addition, the computational complexities of all methods are analyzed and  compared. Numerical simulation show that all proposed methods could approach the CRLB at medium and high SNR regions, SNR$>-5$ dB.
	More importantly, the energy efficiency of the proposed structure is also derived and analyzed. In terms of energy efficiency, the proposed structure is almost identical to the conventional sub-connected HAD, and much higher than fully-digital array.
\end{enumerate}


The remainder of this paper is organized as follows. Section \ref{sec_sys} describes the system model of designed $\mathrm{H}^2$AD MIMO structure. In Section \ref{sec_proposed}, we proposed a DOA framework for the $\mathrm{H}^2$AD structure. Based on that, four high performance estimators are proposed in Section \ref{sec_methods}. In addition, the performance of the proposed structure and methods is analyzed in \ref{sec_perf}. We present our simulation results in Section \ref{sec_simu}. Finally, we draw conclusions in Section \ref{sec_con}.

\emph{Notations:} In this paper, signs $(\cdot)^T$, $(\cdot)^H$, $|\cdot|$ and $\|\cdot\|$ represent transpose, conjugate transpose, modulus and norm, respectively. $\mathbf{x}$ and $\mathbf{X}$ in bold typeface are adopted to represent vectors and matrices, respectively, while scalars are presented in normal typeface, such as $x$. $\mathbf{I}_M$ represents the $M\times M$ identity matrix, $\mathbf{0}_{a\times b}$ denotes the $a\times b$ matrix of all zeros and $\mathbf{1}_M$ denotes the $M\times 1$ vector of all $1$. Furthermore, $\mathbb{E}[\cdot]$, $\mathbb{R}[\cdot]$ and $\mathbb{I}[\cdot]$ denotes the expectation operator, keep the real and image part of a complex-valued number, respectively. $\mathbf{Tr}(\cdot)$  denotes the matrix trace.

\section{System model}\label{sec_sys}

Figure~\ref{fig_sys_mod} shows three different antenna arrays, and Figure \ref{fig: sub_figure1} is the proposed $\mathrm{H}^2$AD structure. Consider a far-field narrowband signal $s(t)e^{j2\pi f_ct}$ impinge on the HAD antenna array, where $s(t)$ is the baseband signal, and $f_c$ is the carrier frequency. We assume a uniform linear array (ULA) with the $M$ antennas that is divided into $Q$ groups, and group $q$ has $K_q$ subarrays with each subarray having $M_q$ antennas. In addition, each subarray contains $M_q$ antennas, i.e.,
\begin{align}
	M =\sum_{q=1}^{Q}N_q=\sum_{q=1}^{Q}K_qM_q,~q\in S_Q=\{1,2,\cdots,Q\}.
\end{align}
In this structure, $M_1\neq M_2 \neq\cdots\neq M_Q$.
Note that, when $M_1=M_2=,\cdots,=M_Q$, as shown in Figure \ref{fig: sub_figure2}, this array can be represented as a common sub-connected hybrid architectures, 
also called a homogeneous structure in this paper.
In other words, the traditional sub-connected HAD array can be considered as a special case of the proposed $\mathrm{H}^2$AD structure. Furthermore, when $M_1=M_2=,\cdots,=M_Q=1$, proposed array will degenerate into a fully-digital array. Therefore, the fully-digital array is also a special case of the sub-connected HAD array. 

In the $q$-th group, the received signal at the $m$-th antenna of the $k$-th subarray is expressed as
\begin{align}
	x_{q,k,m}(t)=s(t)e^{j2\pi (f_ct-f_c\tau_{q,k,m})}+w_{q,k,m}(t),
\end{align}
where $w_{q,k,m}(t)\sim\mathcal{C}\mathcal{N}(0,\sigma^2_w)$ is the additive white Gaussian noise (AWGN), and $\tau_{q,k,m}$ are the propagation delays determined by the direction of the source relative to the array given by
\begin{align}
	\tau_{q,k, m}=\tau_0-\frac{d_m \sin \theta_0}{c},
\end{align}
where $\tau_0$ is the propagation delay from the emitter to a reference point on the array, $c$ is the speed of light, and $d_m$
denotes the distance from a common reference point to the $m$-th antenna of the $k$-th subarray. Since we choose the left of the array as the reference point in this paper, the $d_m=(km-1)d$. Assuming $\phi_{q,k,m}$ be the corresponding phase for analog beamforming corresponding to the $k$-th RF chain and the $m$-th antenna $(m = 1, 2, \cdots , M_q)$, and then the output signal of the $k$-th subarray is
\begin{align}
	y_{q,k}(t)=\frac{1}{\sqrt{M_q}}\sum_{m=1}^{M_q}x_{q,k,m}(t)e^{-j\phi_{q,k,m}}.
\end{align}
Stacking all $K_q$ subarray outputs and passes through parallel RF chains, the baseband signal vector of $q$-th group is formed by
\begin{align} \label{yq}
	\mathbf{y}_q(t)=\mathbf{B}_{A,q}^H\mathbf{a}(\theta_0)s(t)+\mathbf{w}(t),
\end{align}
where $\mathbf{w}(t)=\left[w_1(t), w_2(t), \ldots, w_{N_q}(t)\right]^T\in\mathbb{C}^{N_q \times 1}$ is the AWGN vector, $\mathbf{B}_{A,q}$ is a block diagonal matrix,  whose $k$-th block diagonal element can be represented as
\begin{align}
	\mathbf{b}_{A,q,k}=\left[e^{j\phi_{k,1}},e^{j\phi_{k,2}},\cdots,e^{j\phi_{k,M_q}}\right]^T,
\end{align}
and the vector $\mathbf{a}_q(\theta_0)\in\mathbb{C}^{N_q \times 1}$ is the array manifold defined as
\begin{align}
	\mathbf{a}_q\left(\theta_0\right)=\left[1, e^{j \frac{2 \pi}{\lambda} d \sin \theta_0}, \cdots, e^{j \frac{2 \pi}{\lambda}(N_q-1) d \sin \theta_0}\right]^T.
\end{align}
Via ADC, (\ref{yq}) can be further expressed as
\begin{align}
	\mathbf{y}_q(n)=\mathbf{B}_{A,q}^H\mathbf{a}_q(\theta_0)s(n)+\mathbf{w}(n),
\end{align}
where $n=1,2,\cdots,N$, and $N$ is the number of snapshots.

\section{Proposed high-performance DOA framework for $\mathrm{H}^2$AD structure}\label{sec_proposed}
\begin{figure}
	\centerline{\includegraphics[width=3.5in]{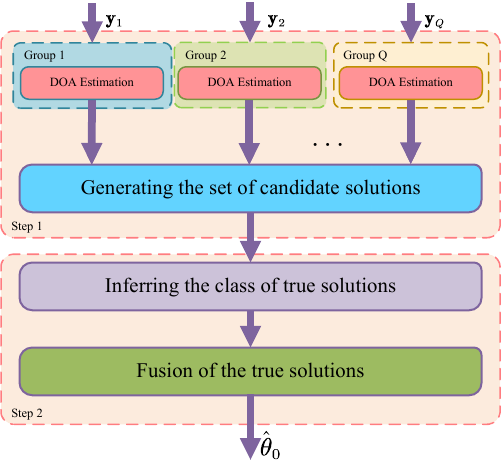}}
	\caption{Proposed a new $\mathrm{H}^2$AD structure for massive MIMO with ULA.\label{fig_alg_flow}}
\end{figure}

In this section, to rapidly eliminate the pseudo solutions and get the correct direction, a high-performance DOA estimator based on $\mathrm{H}^2$AD structure are proposed.
As shown in Figure \ref{fig_alg_flow}, the proposed DOA estimator can be divided into two steps: 1) all subarray groups measure the DOA and generate the set of candidate solutions. 2) infer the clustering of true solutions and merge the true solutions and output the final estimated value.

Assume that the analog beamforming vector $\mathbf{b}_{A,q,k}=[1,1,\cdots,1]^T$, the output vector of $K_q$ subarrays is
\begin{align}
	\mathbf{y}_q(n)=g_q\mathbf{a}_{M_q}(\theta_0)s(n)+\mathbf{w}(n),
\end{align}
where
\begin{align}
	\mathbf{a}_{M_q}\left(\theta_0\right)=\left[1, e^{j \frac{2 \pi}{\lambda} M_q d \sin \theta_0}, \cdots, e^{j \frac{2 \pi}{\lambda}(K_q-1) M_q d \sin \theta_0}\right]^T,
\end{align}
and 
$g_q$ is the gain factor of each subarray, which can be denotes as
\begin{align}
	g_q & =\sum_{m=1}^{M_q} e^{j \frac{2 \pi}{\lambda}(m-1) d \sin \theta_0} \\\nonumber
	& =\frac{1-e^{j \frac{2 \pi}{\lambda} M_q d \sin \theta_0}}{1-e^{j \frac{2 \pi}{\lambda} d \sin \theta_0}}
\end{align}
Then, according to \cite{shuHADDOA2018tcom}, the pseudo spectrum corresponding to the virtual antenna array can be given by
\begin{align} \label{}
	P_{RM}(\theta)=\frac{1}{\left\|\mathbf{U}_N^H \mathbf{a}_{M_q}(\theta)\right\|^2},
\end{align}
which spectral peak corresponds to the desire DOA estimation and the root-MUSIC \cite{1993The} is reliable due to its low-complexity and excellent asymptotic performance. Furthermore, let us define $\mathbf{V}=\mathbf{U}_N\mathbf{U}_N^H$, the polynomial equation can be expressed as
\begin{align} \label{p}
	p(\theta)&=\mathbf{a}_{M_q}^H(\theta)\mathbf{U}_N\mathbf{U}_N^H\mathbf{a}_{M_q}(\theta)\triangleq p(z)\\\nonumber
	&=\sum_{i=1}^{K_q}\sum_{j=1}^{K_q}z^{-(i-1)}\mathbf{V}_{ij}z^{j-1}=0,
\end{align}
where $\mathbf{V}_{ij}$ is the element in the $i$-th column of the $j$-th row of $\mathbf{V}$ and
\begin{align} \label{}
	z=e^{j\xi_q},
\end{align}
where $\xi_q=2\pi M_qdsin\theta/\lambda $. The above polynomial equation (\ref{p}) has $2K_q-2$ roots, i.e., $z_i,i=1,\cdots,2K_q-2$, which implies the existence of multiple emitter directions as follows
\begin{align} \label{}
	\hat{\Theta}_{R M}=\left\{\hat{\theta}_i, i \in\{1,2, \cdots, 2 K_q-2\}\right\},
\end{align}
where
\begin{align} \label{}
	\hat{\theta}_i=\arcsin \left(\frac{\lambda \arg z_i}{2 \pi M_q d}\right).
\end{align}
Then, the angle corresponding to the nearest root to the unit circle is selected as the DOA estimation $\theta_{q}$ of $q$-th array group and
\begin{align} \label{}
	\hat{\xi}_q=\frac{2\pi}{\lambda} M_qdsin\theta_{q}.
\end{align}

Since each virtual antenna corresponds to a subarray, there exists phase ambiguity that needs to be eliminated, which can be expressed as
\begin{align} \label{}
	p(\hat{\xi}_q)=p(\hat{\xi}_q+2\pi j).
\end{align}
Therefore, a feasible set containing $M_q$ solutions as follows
\begin{align} \label{THETA_q}
	\hat{\Theta}_{q}=\left\{\hat{\theta}_{q, j}, j \in\{1,2, \cdots, M_q\}\right\},
\end{align}
where
\begin{align} \label{}
	\hat{\theta}_{q, j}=\arcsin \left(\frac{\lambda\left(\arg (e^{j\hat{\xi}_q})+2 \pi j\right)}{2 \pi M_q d}\right) .
\end{align}
Combing all $Q$ groups gives
\begin{align}\label{angleCandAll}
	\begin{split}
		\left \{
		\begin{array}{ll}
			\frac{2\pi}{\lambda}M_1dsin\theta_{1,j_1}=\hat{\xi}_1+2\pi j_1\\
			\frac{2\pi}{\lambda}M_2dsin\theta_{2,j_2}=\hat{\xi}_2+2\pi j_2\\
			\quad\quad\quad\vdots\quad\quad\quad\quad\quad\quad\quad\vdots\\
			\frac{2\pi}{\lambda}M_Qdsin\theta_{Q,j_Q}=\hat{\xi}_Q+2\pi j_Q
		\end{array}
		\right.
	\end{split},
\end{align}
where $j_q\in\{1,2,\cdots,M_q\}$. Furthermore, the candidate angle set is form as
\begin{align} \label{}
	\hat{\Theta}=\left\{\hat{\Theta}_{1}, \hat{\Theta}_{2} \cdots, \hat{\Theta}_{Q}\right\},
\end{align}
which contains $\sum_{q=1}^QM_q$ solutions and it is a challenging problem to select the true solution from them. Observing (\ref{angleCandAll}), every candidate set contains a true estimated value and $M-1$ pseudo solutions. For the $q$-th subarray group, the true solution and pseudo solution can be respectively expressed as
\begin{align}
	\hat{\theta}_{t,q}=\theta_0+w_q ,
\end{align}
\begin{align}\label{falsetheta}
	\hat{\theta}_{f,q,m}=\theta_0+w_q+\Xi_{q,m},~m=1,2,\cdots,M_q,
\end{align}
where $w_q$ is the estimated error of the $q$-th subarray group and $\Xi_{q,m}$ is a constant determined by $M_q$ and $m$. As the power of noise $w_q$  tends to  $0$, we have
\begin{align}\label{theta_tq}
	\hat{\theta}_{t,1}\approx\hat{\theta}_{t,2}\approx\cdots\approx\hat{\theta}_{t,Q}\approx\theta_0 ,
\end{align}
which forms a class of true solutions, and
\begin{align}
	\hat{\theta}_{f,1,m}\neq\hat{\theta}_{f,2,m}\neq\cdots \neq\hat{\theta}_{f,Q,m}\neq\theta_0 ,
\end{align}
which forms a class of false solutions. 
In general, false solutions are unequal to each other even under the condition that the noise term is close to zero in (\ref{falsetheta}), $\hat{\theta}_{f,q,m}=\theta_0+\Xi_{q,m}$, due to the different values of $\Xi_{q,m}$.
The former means the distance between true solutions goes to zero when the power of noise $w_q$  is far lower than that of useful signal. Conversely, there is no such property for the class of false solutions.

Let us define the class of true solutions as follows 
\begin{equation}
	\hat{\Theta}_{t}=\left\{ \hat{\theta}_{t,1},\hat{\theta}_{t,2},\cdots,\hat{\theta}_{t,Q} \right\},
\end{equation}
where $\hat{\theta}_{t,q}$ denotes the estimated true solution of the $q$ subarray group.
The corresponding class of false solutions can be defined as
\begin{equation}
	\hat{\Theta}_{f}=\hat{\Theta}\cap\hat{\Theta}_t^c.
\end{equation}

\section{Proposed Machine-learning Methods for Deciding the Class of True Solutions}\label{sec_methods}
In the previous section, we have completed Step 1 of our proposed framework, i.e., generating the set of candidate solutions, which should be divided into two classes: false and true. In this section, we will focus on how to infer the class of true solutions and combine all solutions in this class to form an optimal output. To make a high-precision classification, four machine learning (ML) methods will be presented.  The first two methods are based on minimum distance and called WGMD and WLMD while the remaining two ones are to make an enhancement over conventional ML methods, called accelerated local weighted k-means and improved DBSCAN.

\subsection{Proposed Weighted Global Minimum Distance Clustering}

According to (\ref{theta_tq}), all true solutions are close to each other.
Therefore, the extraction of the true angle can be transmitted into finding the same value between all array groups.

The sum distance between solutions of adjacent array groups.
\begin{align} \label{gsProb}
	\hat{\Theta}_{t}=\underset{\hat{\theta}_{RM,j_q}\in\hat{\Theta}_q, q=1,2,\cdots,Q-1}{\arg \min }\sum_{q=1}^{Q-1}\left\|\hat{\theta}_{RM, j_q}-\hat{\theta}_{RM, j_{q+1}}\right\|^2.
\end{align}

The final DOA estimate can be obtained by weighting and combining the Q true solutions
\begin{align} \label{thetaEstConbine}
	\hat{\theta}=\sum_{q=1}^Q w_q \hat{\theta}_{t,q}.
\end{align}
The mean square error (MSE) of $\hat{\theta}$ can be written as 
\begin{align} \label{MSE}
	\mathbb{E}\left[\left(\hat{\theta}-\theta_0\right)^2\right]&=\mathbb{E}\left[\left(\sum_{q=1}^Q w_q \hat{\theta}_{t,q}-\theta_0\right)^2\right] \nonumber\\
	&=\sum_{q=1}^Q w_q^2\mathbb{E}\left[\left(\hat{\theta}_{t,q}-\theta_0\right)^2\right],
\end{align}
where $\mathbb{E}\left[\left(\hat{\theta}_{t,q}-\theta_0\right)^2\right]$ is the MSE of $\hat{\theta}_{t,q}$. 
Without losing generality, we replace the MSE of $\hat{\theta}_{t,q}$ with the corresponding CRLB. Consequently, the problem can be converted into
\begin{align} \label{w_op}
	&\min_{w_q}~~~\sum_{q=1}^Q w^2_q CRLB_q \nonumber\\
	&s.t. ~~~~\sum_{q=1}^Q w_q=1.
\end{align}

From \cite{shuHADDOA2018tcom}, we can achieve the 
the Fisher information matrix (FIM) of the $q$-th HAD group array, which is given by (\ref{FIMHAD}),
\begin{figure*}[!htb]
	\begin{align}\label{FIMHAD}
		FIM_q = \frac{8\pi^2\gamma^2\cos^2\theta_0}{\lambda^2M_a(\gamma N_q\|g_q\|^2+M_a)^2} \left[\|g_q\|^4(N_q\nu-\mu^2)(\gamma N_q\|g_q\|^2 +M_a)+M_aN_q^2\left(\|g_q\|^2\|\Gamma\|^2-\mathbb{R}\left[\left(\Gamma^Hg_q\right)^2\right]\right)\right]
	\end{align}
	\hrulefill
	\vspace*{4pt}
\end{figure*}
where
\begin{align}
	\Gamma = \sum_{m_q=1}^{M_q}\left(m_q-1\right)d e^{j\frac{2\pi}{\lambda}(m_q-1)d},
\end{align}
\begin{align}
	\mu = \frac{1}{2}M_qK_q(K_q-1),
\end{align}
\begin{align}
	\nu = \frac{1}{6}M_q^2K_q(K_q-1)(2K_q-1).
\end{align}
Referring to \cite{T.Engin-09}, the CRLB is given by
\begin{align}
	CRLB = \frac{1}{N}FIM^{-1}.
\end{align}
Therefore, the CRLB of $\hat{\theta}_{t,q}$ expressed as (\ref{CRLBHAD}).
\begin{figure*}[!htb]
	\begin{align}\label{CRLBHAD}
		CRLB_q = \frac{\lambda^2M_a(\gamma N_q\|g_q\|^2+M_a)^2}{8N\pi^2\gamma^2\cos^2\theta_0\left[\|g_q\|^4(N_q\nu-\mu^2)(\gamma N_q\|g_q\|^2 +M_a)+M_aN_q^2\left(\|g_q\|^2\|\Gamma\|^2-\mathbb{R}\left[\left(\Gamma^Hg_q\right)^2\right]\right)\right]} .
	\end{align}
	\hrulefill
	\vspace*{4pt}
\end{figure*}

Next, we can focus on the derivation of (\ref{w_op}).

\textit{Theorem 1:} The closed form expression of the optimization problem is given by
\begin{align} \label{wqFinal}
	w_q=\frac{C R L B_q^{-1}}{\sum_{q=1}^Q\left(C R L B_q\right)^{-1}}.
\end{align}
\textit{Proof:} See Appendix A. $\hfill\blacksquare$

In general, the proposed DOA algorithm are divided into three steps: 1) perform DOA estimation for all group arrays. 2) eliminate phase ambiguity. 3) integrate all solutions of all group arrays. This method could be named as WGMD.
The overall algorithm is shown in Algorithm \ref{alg:GS}.

\begin{algorithm}[H]
	\caption{Proposed WGMD.}\label{alg:GS}
	\begin{algorithmic}
		\STATE 
		\STATE {\textbf{Input:}}$~\mathbf{y}(n),~ n=1,2,\cdots,N.$
		\STATE \hspace{0.5cm} \textbf{Initialization:} ~divide $\mathbf{y}(n)$ into $\mathbf{y}_q(n)$, $q=1,2,\cdots,Q$. Calculate the $CRLB_q$.
		\STATE \hspace{0.5cm} \textbf{for} $q=1,2,\cdots,Q$ \textbf{do},
		\STATE \hspace{1cm} perform the root-MUSIC method for $\mathbf{y}_q(n)$, and obtain the solution set, $\hat{\Theta}_{q}$.
		\STATE \hspace{0.5cm} \textbf{end for}
		\STATE \hspace{0.5cm} Solve the optimization problem (\ref{gsProb}) to achieve the set $\hat{\Theta}_{t}$.
		\STATE \hspace{0.5cm} substitute $\hat{\Theta}_{t}$ and (\ref{wqFinal}) into (\ref{thetaEstConbine}) to obtain $\hat{\theta}$.
		\STATE {\textbf{Output:}} $\hat{\theta}$.
	\end{algorithmic}
\end{algorithm}

\subsection{Proposed Weighted Local Minimum Distance Clustering}

To reduce the computational complexity of the global search method, we proposed a local search method.
The specific selection process is as follows. The elements in the $q$ candidate set are separated from all the elements in the $q+1$ candidate set by a relative distance, and the two elements with the smallest distance are the true solutions of the two candidate sets, which can be expressed as
\begin{align} \label{localSearch}
	&\left\{\hat{\theta}_{t,q}, \hat{\theta}_{t,q+1}\right\}\nonumber\\
	&~~~~=\underset{\hat{\theta}_{RM,j_q}\in\hat{\Theta}_q, \hat{\theta}_{RM, j_{q+1}}\in\hat{\Theta}_{q+1}}{\arg \min }\left\|\hat{\theta}_{RM, j_q}-\hat{\theta}_{RM, j_{q+1}}\right\|^2,
\end{align}
where $\hat{\theta}_{t,q}$ and $\hat{\theta}_{t,q+1}$ are the true solutions given by the $q$ and $q+1$ groups, respectively, and $\hat{\theta}_{RM,j_q}$ and $\hat{\theta}_{RM,j_{q+1}}$ are the angles in the candidate solution sets of the $q$ and $q+1$ groups. When $Q$ is an even number, the $Q$ groups subarrays can be divided into $Q/2$ sets to compute (\ref{localSearch}), where $q=1,3,\cdots,Q-1$. If $Q$ is an odd number, array groups are divided into $Q/2-1$ sets with one set has three group arrays. This method is named WLMD.


\subsection{Accelerated Local Weighted K-means Clustering}
Inspired by the k-means algorithm, we propose a low-complexity recursive search method.

When the true solution of the $q$-th group arrays is obtained, denoted by $\hat{\theta}_{q}$, the true solution of $(q+1)$-th group array is given by searching the angle with smallest distance to the $\hat{\theta}_{q}$.
\begin{align} \label{pLR}
	\hat{\theta}_{t,q+1}=\underset{ \hat{\theta}_{RM, j_{q+1}}\in\hat{\Theta}_{q+1}}{\arg \min }\left\|\hat{\theta}_{q}-\hat{\theta}_{RM, j_{q+1}}\right\|^2,
\end{align}
where 
\begin{align} \label{LRre}
	\hat{\theta}_{q+1}=\frac{w_{q}'}{w_{q+1}'}\hat{\theta}_{q} +\frac{w_{q+1}}{w_{q+1}'} \hat{\theta}_{t,q+1},
\end{align}
\begin{align} \label{}
	w_{q+1}' =  \sum_{q'=1}^{q+1}w_{q'}= w_{q}'+w_{q+1},
\end{align}
where $\hat{\theta}_q$ is the estimated result of $q$-th iteration, where $q=2,3,\cdots,Q-1$. 

The initial solution, $\hat{\theta}_2$, is given by
\begin{align} \label{thetaIni}
	\hat{\theta}_2 = \frac{w_{1}}{w_{1}+w_{2}}\hat{\theta}_{t,1} +\frac{w_{2}}{w_{1}+w_{2}} \hat{\theta}_{t,2},
\end{align}
where $\hat{\theta}_{t,1}$ and $\hat{\theta}_{t,1}$ are generated by (\ref{localSearch}). In addition, to minimize the computational complexity of generating a initial solution, two group arrays with smallest $M_q$ are chosen. This method is called ALW-K-means. Details are shown in Algorithm \ref{alg:RS}.

\begin{algorithm}[H]
	\caption{ALW-K-means.}\label{alg:RS}
	\begin{algorithmic}
		\STATE 
		\STATE {\textbf{Input:}}$~\mathbf{y}(n),~ n=1,2,\cdots,N.$
		\STATE \hspace{0.5cm} \textbf{Initialization:} ~divide $\mathbf{y}(n)$ into $\mathbf{y}_q(n)$, $q=1,2,\cdots,Q$. Calculate the $CRLB_q$.
		\STATE \hspace{0.5cm} \textbf{for} $q=1,2,\cdots,Q$ \textbf{do},
		\STATE \hspace{1cm} perform the root-MUSIC method for $\mathbf{y}_q(n)$, and obtain the solution set, $\hat{\Theta}_{q}$.
		\STATE \hspace{0.5cm} \textbf{end for}
		\STATE \hspace{0.5cm} Solve the optimization problem (\ref{pLR}) by (\ref{LRre})-(\ref{thetaIni}) conduct to obtain the set $\hat{\Theta}_{t}$.
		\STATE \hspace{0.5cm} substitute $\hat{\Theta}_{t}$ and (\ref{wqFinal}) into (\ref{thetaEstConbine}) to obtain $\hat{\theta}$.
		\STATE {\textbf{Output:}} $\hat{\theta}$.
	\end{algorithmic}
\end{algorithm}

\begin{figure*}[!htp]
	\centerline{\includegraphics[width=7in]{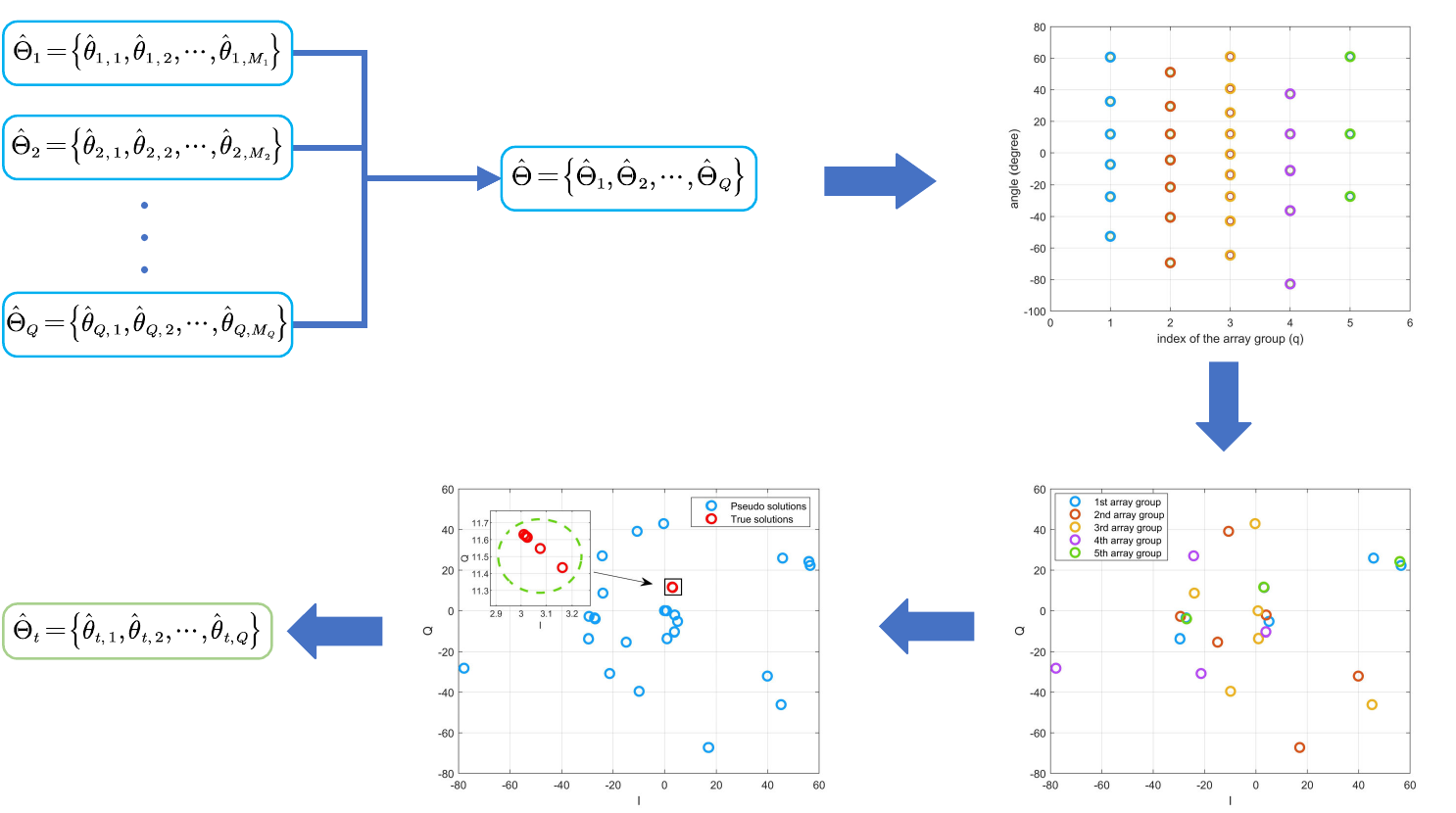}}
	\caption{The framework of the DBSCAN-based search method}\label{fig_cluster}
\end{figure*}

\subsection{Improved Density-based Clustering}
\begin{algorithm}[t]
	\caption{Proposed Improved-DBSCAN.}\label{alg:cluster}
	\begin{algorithmic}
		\STATE 
		\STATE {\textbf{Input:}}$~\mathbf{y}(n),~ n=1,2,\cdots,N.$
		\STATE \hspace{0.5cm} \textbf{Initialization:} ~divide $\mathbf{y}(n)$ into $\mathbf{y}_q(n)$, $q=1,2,\cdots,Q$. Calculate the $CRLB_q$. $Q_c=Q$. $d_d=0$. $\hat{\Theta}_t=\{\}$. maxCount.
		\STATE \hspace{0.5cm} \textbf{for} $q=1,2,\cdots,Q$ \textbf{do},
		\STATE \hspace{1cm} perform the root-MUSIC method for $\mathbf{y}_q(n)$, and obtain the candidate solution set.
		\STATE \hspace{0.5cm} \textbf{end for}
		\STATE \hspace{0.5cm} generate $\hat{\Theta}_c$ by calculating $|\theta|e^{j\theta}$ for candidate solution set.
		\STATE \hspace{0.5cm} Calculate $d_{max}$ by selecting the minimum Euclidean distance of points in $\hat{\Theta}_{c}$, which has the most candidate solutions. 
		\STATE \hspace{0.5cm} $d_u = d_{max}$.
		\STATE \hspace{0.5cm} \textbf{while} Count $<$ maxCount
		\STATE \hspace{1cm} $d_\epsilon = (d_u + d_d) / 2$.
		\STATE \hspace{1cm} Do DBSCAN with $(Q_c,d_\epsilon)$ and achieve the result set $\hat{\Theta}_{db}$, which has the most points.
		\STATE \hspace{1cm} \textbf{if} $length(\hat{\Theta}_{db})==Q$ and all solutions are from different candidate solution sets.
		\STATE \hspace{1.5cm} $\hat{\Theta}_{t}=\hat{\Theta}_{db}$.
		\STATE \hspace{1.5cm} \textbf{break}.
		\STATE \hspace{1cm} \textbf{elseif} $length(\hat{\Theta}_{db})>Q$
		\STATE \hspace{1.5cm} $d_u = d_\epsilon$.
		\STATE \hspace{1cm} \textbf{elseif} $length(\hat{\Theta}_{db})<Q$
		\STATE \hspace{1.5cm} $d_d = d_\epsilon$.
		\STATE \hspace{1cm} \textbf{else}
		\STATE \hspace{1.5cm} \textbf{break}.
		\STATE \hspace{1cm} \textbf{end if}
		\STATE \hspace{1cm} Count = Count + 1.
		\STATE \hspace{0.5cm} \textbf{end while}
		\STATE \hspace{0.5cm} \textbf{if} $length(\hat{\Theta}_{t})>0$
		\STATE \hspace{1cm} substitute $\hat{\Theta}_{t}$ and (\ref{wqFinal}) into (\ref{thetaEstConbine}) to obtain $\hat{\theta}$.
		\STATE \hspace{0.5cm} \textbf{else}
		\STATE \hspace{1cm} $\hat{\theta}=$Nan.
		\STATE \hspace{0.5cm} \textbf{endif}
		\STATE {\textbf{Output:}} $\hat{\theta}$.
	\end{algorithmic}
\end{algorithm}

Since true solutions are gathered and pseudo solutions are scattered, we consider adopting the density-based cluster method.

As shown in Figure \ref{fig_cluster}, all angles of all group arrays are obtained firstly. Then,
convert the one-dimensional information $\theta$ into a two-dimensional scattered point diagram by $|\theta|e^{j\theta}$ to generate a new candidate solution set $\hat{\Theta}_c$
\begin{align} \label{}
	&I(\theta) = \mathbb{R}\left[\theta|e^{j\theta}|\right], \nonumber\\
	&Q(\theta) = \mathbb{I}\left[\theta|e^{j\theta}|\right].
\end{align}

Observing the two-dimensional figure, the $Q$ points corresponding to true angles are almost coincident and others are discrete. Therefore, the DBSCAN method is able to be adopted \cite{ester1996density}. DBSCAN algorithm has two parameters, $Q_c$ and $d_\epsilon$. The algorithm can be summarized as three steps: 1) select a point randomly and search the points within a Euclidean distance of $d_\epsilon$ 2) if the number of points that meet the conditions is over or equal the $Q_c$, these points are considered as one cluster. Then search other points based on points above. 3) repeat steps 1) and 2) for points that are not traversed.

However, in our work, only $Q$ nearest point need to be distinguished. As a result, the minimum number of reachable points, $Q_c$, can be set as $Q$. In addition, to ensure that the points belonging to the same solution set would not be clustered as one group directly, the maximum value of $d_\epsilon$ should be set as the minimum distance of points in the candidate set $\hat{\Theta}_{q}$, which has the most candidate solutions. The specific value could be chosen by the dichotomy method. Details of the whole algorithm are shown in Algorithm \ref{alg:cluster}, which is named improved-DBSCAN.

\section{Theoretical Analysis}\label{sec_perf}
In this section, we analyze the theoretical performance of the proposed structure and the computational complexity of proposed methods.

\subsection{Performance Accuracy}
CRLB is a low bound of the error variance for an unbiased estimator. Thus, we derive the closed-form expression of the CRLB corresponding to the proposed $\mathrm{H}^2$AD structure. Furthermore, we adopt it as the benchmark of proposed structure. 

\textit{Theorem 2:} The lower bound of variance for unbiased DOA estimation method with the $\mathrm{H}^2$AD structure is given by
\begin{align}\label{CRLB_sigma}
	\sigma_{\theta_0}^2 \geq \frac{1}{N}\mathbf{FIM}^{-1},
\end{align}
where
\begin{align}
	\mathbf{FIM} = \sum_{q=1}^{Q}\mathbf{FIM}_{q}.
\end{align}
Then, the closed-form expression of CRLB is given by (\ref{CRLB_HHAD}).
\begin{figure*}[!htb]
	\begin{align}\label{CRLB_HHAD}
		CRLB =\left( N\sum_{q=1}^{Q}\frac{8\pi^2\gamma^2\cos^2\theta_0\left[\|g_q\|^4(N_q\nu-\mu^2)(\gamma N_q\|g_q\|^2 +M_a)+M_aN_q^2\left(\|g_q\|^2\|\Gamma\|^2-\mathbb{R}\left[\left(\Gamma^Hg_q\right)^2\right]\right)\right]}{\lambda^2M_a(\gamma N_q\|g_q\|^2+M_a)^2} \right)^{-1}
	\end{align}
	\hrulefill
	\vspace*{4pt}
\end{figure*}

\textit{Proof:} See Appendix B. $\hfill\blacksquare$

\subsection{MSE}
In this subsection, we will derive the MSE of the proposed framework. Assumed that the DOA estimation of all group could achieve the CRLB.
And, all true solutions can be inferred correctly. Referring to (\ref{MSE}), (\ref{w_op}) and (\ref{wqFinal}), the MSE of the proposed methods is
\begin{align} 
	MSE &= \sum_{q=1}^Q w^2_q CRLB_q \nonumber\\
	&= \sum_{q=1}^{Q}\left(\frac{C R L B_q^{-1}}{\sum_{q=1}^Q\left(C R L B_q\right)^{-1}}\right)^2 CRLB_q \nonumber\\
	& = \frac{1}{\left(\sum_{q=1}^Q\left(C R L B_q\right)^{-1}\right)^2}\sum_{q=1}^Q\frac{1}{CRLB_q}.
\end{align}
Since $C R L B_q = \frac{1}{N\cdot \mathbf{FIM}_q}$, we have
\begin{align} 
	MSE &= \frac{1}{\left(\sum_{q=1}^Q\left(C R L B_q\right)^{-1}\right)^2}\sum_{q=1}^Q\frac{1}{CRLB_q} \nonumber\\
	&=  \frac{1}{N^2\left(\sum_{q=1}^Q \mathbf{FIM}_q\right)^2}\cdot N\sum_{q=1}^Q \mathbf{FIM}_q\nonumber\\
	& = \frac{1}{N}\left(\sum_{q=1}^Q \mathbf{FIM}_q\right)^{-1}.
\end{align}
It is obvious that the MSE is the same as (\ref{CRLB_sigma}). Thus, the proposed framework could achieve the corresponding CRLB.

\subsection{Computational Complexity}
\begin{table*}
	\footnotesize
	\centering
	\caption{Computational Complexity of Proposed Methods}
	\label{tab1}
	\scalebox{1.3}{
		\begin{tabular}{c|c}
			\hline
			\hline
			Algorithms     	& Complexity          \\
			\hline
			Poposed WGMD & $\mathcal{O}(\sum_{q=1}^{Q}K_q^2N+(2K_q-2)^2(2K_q+NK_q-2)+\prod_{q=1}^{Q}M_q+Q)$ \\
			\hline
			Proposed WLMD	&  $\mathcal{O}(\sum_{q=1}^{Q}K_q^2N+(2K_q-2)^2(2K_q+NK_q-2)+\sum_{q=1}^{Q/2}M_{2q-1}M_{2q}+Q)$   \\
			\hline
			ALW-K-means		& $\mathcal{O}(\sum_{q=1}^{Q}K_q^2N+(2K_q-2)^2(2K_q+NK_q-2)+\sum_{q=3}^{Q}M_q+ 2Q-4)$   \\
			\hline
			Improved-DBSCAN 	& $\mathcal{O}(\sum_{q=1}^{Q}K_q^2N+(2K_q-2)^2(2K_q+NK_q-2)+(\sum_{q=1}^{Q}M_q)\log(\sum_{q=1}^{Q}M_q)+Q)$     \\
			\hline
			\hline
	\end{tabular}}
\end{table*}

In this section, we investigate the computational complexity of proposed methods.
As shown in Figure \ref{fig_alg_flow}, all proposed methods have three steps and only the second step is different except that proposed ALW-K-means integrated step 3 into step 2. Thus, the computational complexity of the first and last step is the same. The first step is to calculate root-MUSIC methods for all subarray groups, whose computational complexity is $\mathcal{O}\left(\sum_{q=1}^{Q}K_q^2N+(2K_q-2)^2(2K_q+NK_q-2)\right)$. Step 3 is a linear weight method.
The weighting factors could be calculated in advance. Thus, the computational complexity is $\mathcal{O}(Q)$. Then, let us analyze step 2. the computational complexity of the globe search and local search is $\mathcal{O}(\prod_{q=1}^{Q}M_q)$ and $\mathcal{O}(\sum_{q=1}^{Q/2}M_{2q-1}M_{2q})$, respectively. 
And, the computational complexity of the DBSCAN is given by $\mathcal{O}(\sum_{q=1}^{Q}M_q\log(\sum_{q=1}^{Q}M_q))$. The overall computational complexity of steps 2 and 3 for the ALW-K-means is expressed as $\mathcal{O}(M_1 M_2+\sum_{q=3}^{Q}M_q+ 2Q-4 )$.
So far, the computational complexity of all steps for all proposed methods has been investigated.
The computational complexity of all proposed algorithms are shown in Table \ref{tab1}.

%

\section{Simulation Results}\label{sec_simu}

In this section, we present simulation results to assess the performance of proposed DOA estimator and the corresponding CRLB as a performance benchmark. Assuming a source impinge on the array with $\theta_0=41^{\circ}$, $Q=3$, $N=100$, $K_1=K_2=K_3=16$, $ M_1=7 $, $ M_2=11 $ and $ M_3=13 $. In our simulation, all results are averaged over 5000 Monte Carlo realizations and root-mean-squared error (RMSE) is employed to indicate the performance, which is given by
\begin{align} \label{}
	RMSE = \left(\frac{1}{N_{simu}}\sum_{n_{simu}}^{N_{simu}}(\hat{\theta}_{n_{simu}}-\theta_0)^2\right)^{\frac{1}{2}},
\end{align}
where $N_{simu}$ is the number of Monte Carlo realizations.

\begin{figure}[!htb]
	\centerline{\includegraphics[width=3.5in]{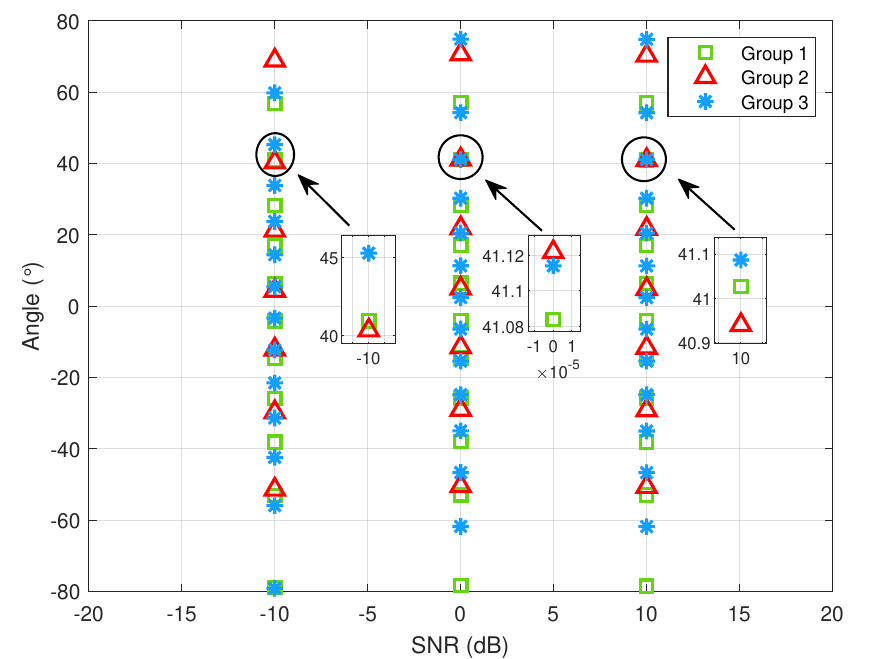}}
	\caption{Angle of feasible solution set versus SNR of the proposed method}\label{figure2}
\end{figure}

Figure \ref{figure2} plots the angle of feasible solution set for each group versus SNR of the proposed method for $ M_1=7 $, $ M_2=11 $ and $ M_3=13 $. From Figure \ref{figure2}, it can be seen that there are $Q$ points in the candidate solution set generated by the $Q$ groups that are closest to each other, and these points are considered as $Q$ true solutions. And, other points are scattered in other space. In addition, as SNR increases, the distance between true solutions becomes shorter.

\begin{figure}[!htb]
	\centerline{\includegraphics[width=3.5in]{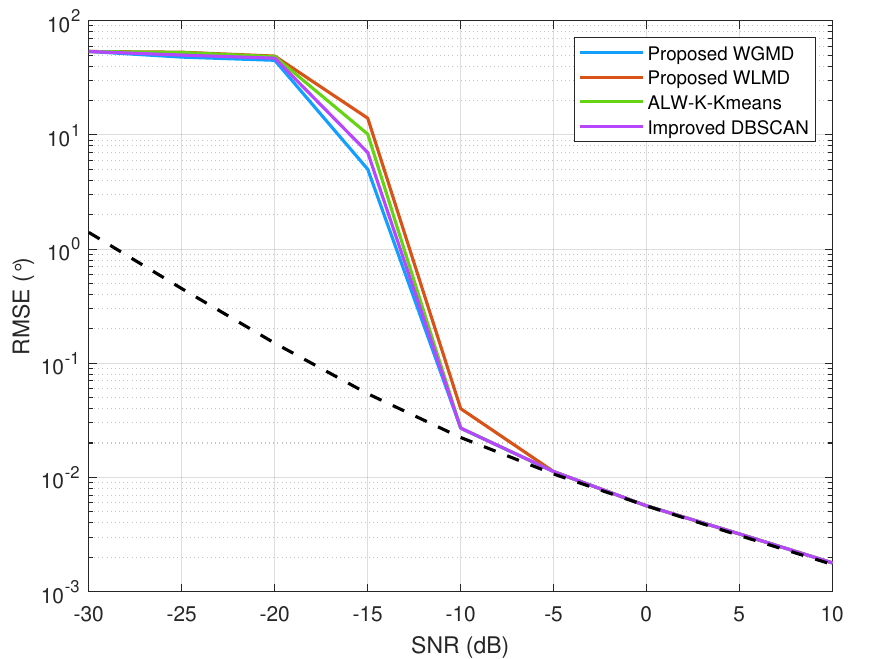}}
	\caption{RMSE versus SNR of the proposed method}\label{fig_RMSE}
\end{figure}

To further validate the conclusion of Figure \ref{figure2}, Figure \ref{fig_RMSE} plots the root mean squared error (RMSE) versus SNR of the proposed DOA methods, where the corresponding CRLB is also plotted as a performance benchmark. 
From Figure \ref{fig_RMSE}, it is seen that the proposed methods are able to achieve the corresponding CRLB when SNR$>-5$ dB.
Since all search methods could find out true solutions with a $100\%$ accuracy when SNR$>-5dB$, all methods have the same performance and the corresponding curves overlap together. The difference of RMSE between $-20$ dB to $-5$ dB is caused by the different accuracy of different methods. When SNR drops below -10 dB, root-MUSIC fails to generate the right candidate solutions.

\begin{figure}[!htb]
	\centerline{\includegraphics[width=3.5in]{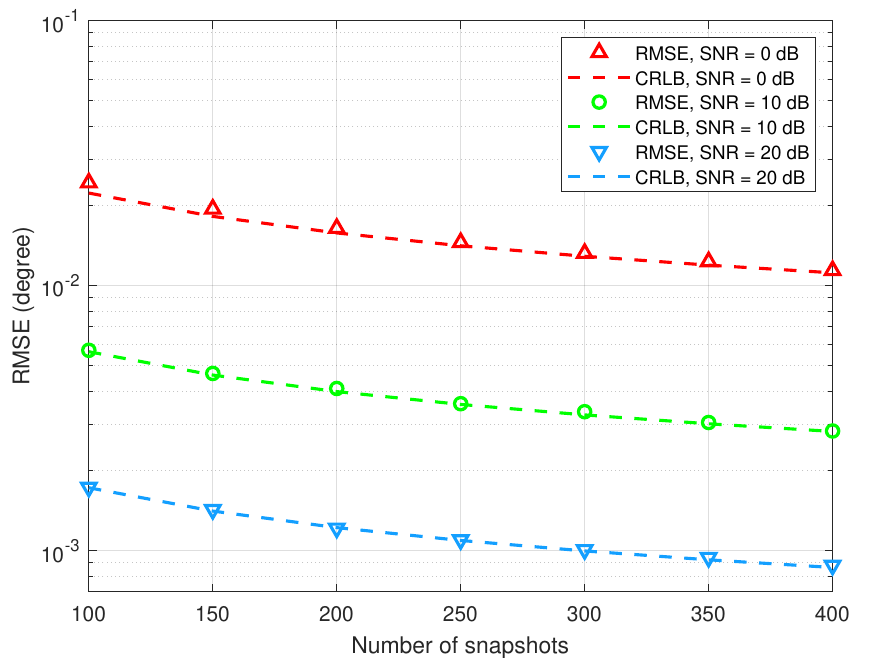}}
	\caption{RMSE versus number of snapshots for the proposed method with $SNR\in\{0, 10, 20\}$ dB}\label{fig_snapshots}
\end{figure}

Figure \ref{fig_snapshots} illustrate the RMSE versus the number of snapshots for the proposed method. The number of snapshots increases from 100 to 400. 
It is obvious that all points of RMSE could approach the corresponding CRLB with three different SNR. In essence, increase of snapshots is the increase of the SNR. Thus, the performance could approach the CRLB, which is consistent with conclusion in Figure \ref{fig_RMSE}.

\begin{figure}[t]
	\centerline{\includegraphics[width=3.5in]{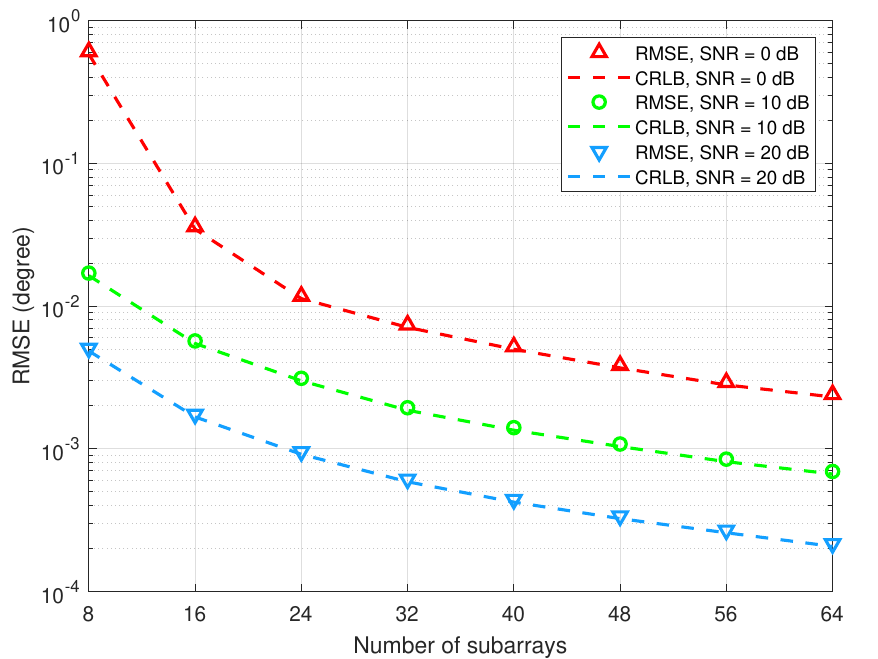}}
	\caption{RMSE versus number of subarrays for the proposed method}\label{fig_an}
\end{figure}

Figure \ref{fig_an} shows the RMSE over the number of subarrays. In this simulation, the value of $M_q$ keeps constant and $K_q$ is increased. Note that all subarray groups have the same subarrays, i.e. $K_1=K_2=\cdots=K_Q$. Simulation values also could reach the CRLB as $K_q$ increases. Combined with Figure \ref{fig_RMSE} and \ref{fig_snapshots}, simulation results show that proposed methods can always achieve a high performance at medium and high SNR.

\begin{figure}[t]
	\centerline{\includegraphics[width=3.5in]{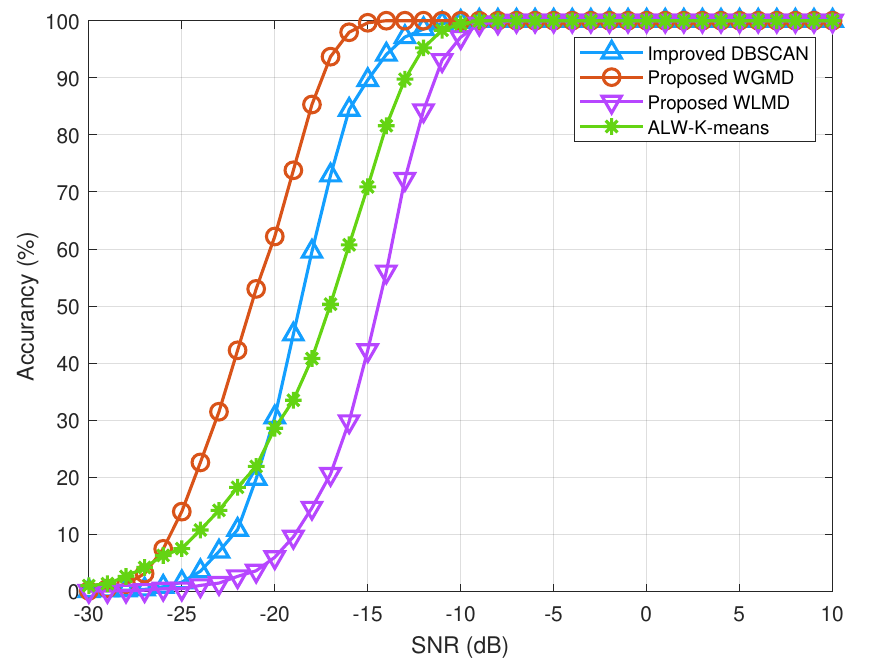}}
	\caption{The accuracy rate of finding out the true solutions over the SNR for the proposed methods}\label{fig_acc}
\end{figure}

The accuracy rate of searching out all true solutions with different proposed methods are shown in Figure \ref{fig_acc}. To better distinguish the performance of proposed algorithms, we change the parameters as: $M=[18,19,20,21]$, $SNR=-30\sim10$ dB. It can be seen that the WGMD has the best performance. Two machine learning based methods have the similar accuracy and are all higher than the WLMD and ALW-K-means. Consider that the WGMD also has the highest complexity, ALW-K-means and improved-DBSCAN are more suitable to be used in practical. Furthermore, the accuracy of Improved-DBSCAN is higher than ALW-K-means when SNR$\in [-20,10]$, but the results is reversed when SNR$\in [-30,20]$. 

\begin{figure}[t]
	\centerline{\includegraphics[width=3.5in]{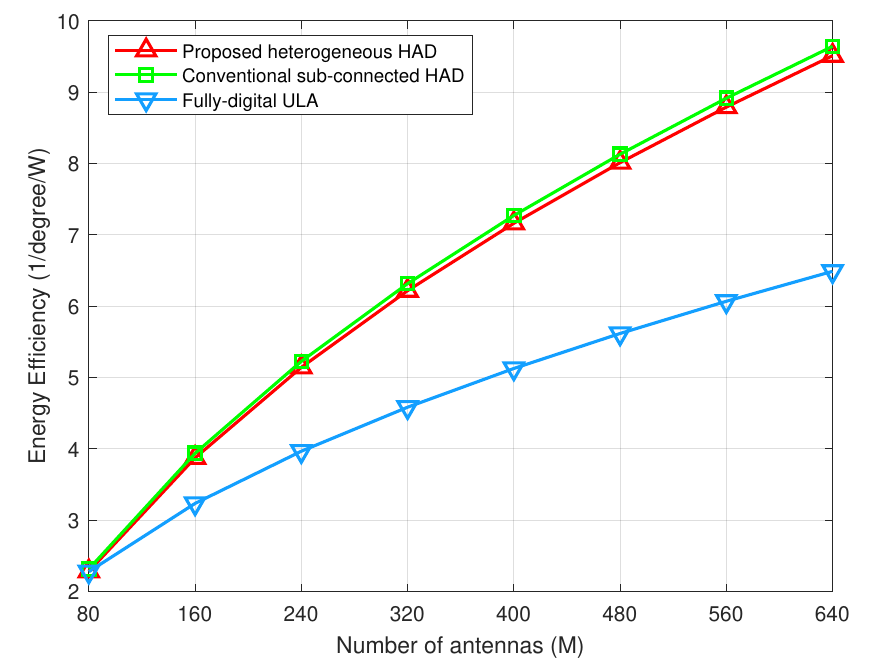}}
	\caption{Energy efficiency versus the number of antennas for different arrays}\label{fig_EE}
\end{figure}

Energy efficiency is also a import factor that assess the proposed antenna array. Thus, the energy efficiency for different array structures is plotted in Figure \ref{fig_EE}.
Referring to \cite{shiDOA2023ojcs}, the energy efficiency factor can be given by
\begin{equation}\label{epsilon}
	\eta_{EE}=\frac{\emph{CRLB}^{-\frac{1}{2}}}{P_{all}}~~1/\mathrm{degree}/\mathrm{W},
\end{equation}
where $P_{all}$ is the total power consumption in the array. It is subject to the number of RF chains, ADCs, low noise amplifiers and other RF devices. The values of parameters are chosen as the same as \cite{shiDOA2023ojcs}. Observing the figure, we find that the proposed $\mathrm{H}^2$AD structure has the similar energy with the conventional sub-connected array and is much higher than the fully-digital ULA. It is worth noting that the advantage of the proposed array will continue to grow as SNR increases.

\section{Conclusions}\label{sec_con}
In this  paper,  to address the  problem of low time-efficiency of removing the phase ambiguity in a conventional sub-connect antenna array, a new general $\mathrm{H}^2$AD MIMO array is designed  and composed of several groups with each group actually being a conventional sub-connect array.  Here, each group is equipped with several sub-connect subarrays with identical number of antennas but the number of antennas of sub-array varies from group to group. The main advantage of the proposed structure has an inner power of rapidly eliminating phase ambiguity within one single sampling period while the traditional sub-connect array usually needs at least two sampling ones. This accelerates the process of DOA measurement and significantly saves the computational time-delay. Then, a two-step framework of DOA measurement was developed. 
Based on that framework, four high-performance DOA estimators were presented. As the number of antennas tends to medium-scale or large-scale, the proposed methods can rapidly eliminate phase ambiguity and achieve excellent estimation performance at almost all SNR except for extremely low SNR. 
In addition, the proposed ML methods could achieve a satisfactory performance with much lower complexity.
The proposed four methods have an accuracy order as follows: WGMD, improved DBSCAN, ALW-K-means and WLMD.
And, the complexity order is presented as follows: ALW-K-means, improved DBSCAN, WLMD, improved DBSCAN and WGMD. In general, we recommend the use of ALW-K-means and improved DBSCAN at most applications, which could achieve a better trade-off between performance and complexity.
In summary, using only single sample period to remove the phase ambiguity, the proposed $\mathrm{H}^2$AD MIMO structure shows a much lower latency than  conventional HAD one,  and  a higher energy efficiency, and lower RF circuit cost and computational complexity  than conventional FD one. Therefore, the proposed $\mathrm{H}^2$AD structure will be very suitable for the future low-latency green wireless networks such as ultra reliable and low latency communications, and integrated sensing and communication.

\section*{APPENDIX A: Proof of Theorem 1}
According to (\ref{w_op}), we can define the Lagrangian function, $L(w_q,\lambda')$, as
\begin{equation}\label{lagFunc}
	L(w_q,\lambda')=\sum_{q=1}^{Q}w_q^2CRLB_q- \lambda'\left(\sum_{q=1}^{Q}w_q-1\right),
\end{equation}
where $\lambda'$ is the Lagrange multiplier. The corresponding partial derivatives are given by 
\begin{equation}
	\frac{\partial L(w_q,\lambda')}{\partial w_q}= 2 w_q CRLB_q-\lambda',
\end{equation}
\begin{equation}
	\frac{\partial L(w_q,\lambda')}{\partial \lambda'}=- \sum_{q=1}^{Q}w_q+1.
\end{equation}
Since (\ref{lagFunc}) is a convex quadratic function of $w_q$, we can set
\begin{equation}
	\begin{split}
		\left \{
		\begin{array}{ll}
			\frac{\partial L(w_q,\lambda')}{\partial w_q} = 0\\
			\frac{\partial L(w_q,\lambda')}{\partial \lambda'}= 0
		\end{array}
		\right.,
	\end{split}
\end{equation}
which yields
\begin{equation}
	w_q=\frac{C R L B_q^{-1}}{\sum_{q=1}^Q\left(C R L B_q\right)^{-1}}.
\end{equation}
Therefore, the proof of Theorem 2 is completed.

\section*{APPENDIX B: FIM for the $\mathrm{H}^2$AD structure}\label{AppB}
In this section, we derive the FIM for the $\mathrm{H}^2$AD structure. The FIM can be expressed as
\begin{equation}
	\mathbf{FIM}_{\mathbf{y}} = \mathbf{Tr}\left\{
	\mathbf{R}_{\mathbf{y}}^{-1} \frac{\partial \mathbf{R}_{\mathbf{y}}}{\partial \theta}\mathbf{R}_{\mathbf{y}}^{-1} \frac{\partial \mathbf{R}_{\mathbf{y}}}{\partial \theta}
	\right\},
\end{equation}
where
\begin{equation}
	\mathbf{y} = [\mathbf{y}_1,\mathbf{y}_2,\cdots,\mathbf{y}_Q]^T =  \mathbf{B}_A^H \mathbf{a}s+\mathbf{w},
\end{equation}
where 
\begin{equation}\label{BA_sub}
	\mathbf{B}_A = \left[
	\begin{array}{cccc}
		\mathbf{B}_{A,1} & \mathbf{0} & \cdots & \mathbf{0} \\
		\mathbf{0} & \mathbf{B}_{A,2} & \cdots & \mathbf{0} \\
		\vdots & \vdots & \ddots & \vdots \\
		\mathbf{0} & \mathbf{0} & \cdots & \mathbf{B}_{A,Q}
	\end{array}
	\right],
\end{equation}
\begin{equation}
	\mathbf{a} = \left[\varphi_1\mathbf{a}_1^T,\varphi_2\mathbf{a}_2^T,\cdots,\varphi_Q\mathbf{a}_Q^T\right]^T,
\end{equation}
where
\begin{equation}\label{varphiq}
	\varphi_q = \left\{
	\begin{array}{ll}
		1 &,q=1 \\
		e^{j \frac{2 \pi}{\lambda} d \sin \theta_0\sum_{q_j=1}^{q-1}N_{q_j} }  &,q>1
	\end{array}
	\right.,
\end{equation}
\begin{equation}\label{Ry}
	\mathbf{R}_{\mathbf{y}} = \gamma \mathbf{B}_A^H \mathbf{a}\mathbf{a}^H \mathbf{B}_A + \mathbf{I}.
\end{equation}
Then, submitting (\ref{BA_sub})-(\ref{varphiq}) into (\ref{Ry}), we can simplify the (\ref{Ry}) as (\ref{Rysub}),
\begin{figure*}[!htb]
	\begin{align}\label{Rysub}
		\mathbf{R}_{\mathbf{y}} &=\gamma
		\left[
		\begin{array}{cccc}
			\mathbf{B}_{A,1}^H & \mathbf{0} & \cdots & \mathbf{0} \\
			\mathbf{0} & \mathbf{B}_{A,2}^H & \cdots & \mathbf{0} \\
			\vdots & \vdots & \ddots & \vdots \\
			\mathbf{0} & \mathbf{0} & \cdots & \mathbf{B}_{A,Q}^H
		\end{array}
		\right]\left[
		\begin{array}{c}
			\varphi_1\mathbf{a}_1  \\
			\varphi_2\mathbf{a}_2  \\
			\vdots  \\
			\varphi_Q\mathbf{a}_Q
		\end{array}
		\right]\left[\varphi_1^H\mathbf{a}_1^H,\varphi_2^H\mathbf{a}_2^H, \cdots,\varphi_Q^H\mathbf{a}_Q^H\right]
		\left[
		\begin{array}{cccc}
			\mathbf{B}_{A,1} & \mathbf{0} & \cdots & \mathbf{0} \\
			\mathbf{0} & \mathbf{B}_{A,2} & \cdots & \mathbf{0} \\
			\vdots & \vdots & \ddots & \vdots \\
			\mathbf{0} & \mathbf{0} & \cdots & \mathbf{B}_{A,Q}
		\end{array}
		\right]+\mathbf{I} \nonumber\\
		&=\gamma
		\left[
		\begin{array}{cccc}
			\mathbf{B}_{A,1}^H\mathbf{a}_1\mathbf{a}_1^H\mathbf{B}_{A,1} & \mathbf{0} & \cdots & \mathbf{0} \\
			\mathbf{0} & \mathbf{B}_{A,2}^H\mathbf{a}_2\mathbf{a}_2^H\mathbf{B}_{A,2} & \cdots & \mathbf{0} \\
			\vdots & \vdots & \ddots & \vdots \\
			\mathbf{0} & \mathbf{0} & \cdots & \mathbf{B}_{A,Q}^H\mathbf{a}_Q\mathbf{a}_Q^H\mathbf{B}_{A,Q}
		\end{array}
		\right]+\mathbf{I}
		=
		\left[
		\begin{array}{cccc}
			\mathbf{R}_{\mathbf{y}_1} & \mathbf{0} & \cdots & \mathbf{0} \\
			\mathbf{0} & \mathbf{R}_{\mathbf{y}_2} & \cdots & \mathbf{0} \\
			\vdots & \vdots & \ddots & \vdots \\
			\mathbf{0} & \mathbf{0} & \cdots & \mathbf{R}_{\mathbf{y}_Q}
		\end{array}
		\right]
	\end{align}
	\hrulefill
	\vspace*{4pt}
\end{figure*}
where
\begin{align}
	\mathbf{R}_{\mathbf{y}_q} = \gamma\mathbf{B}_{A,q}^H\mathbf{a}_q\mathbf{a}_q^H\mathbf{B}_{A,q} + \mathbf{I}.
\end{align}
Thus, the FIM can be given by
\begin{figure*}[!htb]
	\begin{align}
		\mathbf{FIM}_{\mathbf{y}} &=\mathbf{Tr}\left\{ \left[
		\begin{array}{cccc}
			\mathbf{R}_{\mathbf{y}_1}^{-1} \frac{\partial \mathbf{R}_{\mathbf{y}_1}}{\partial \theta}\mathbf{R}_{\mathbf{y}_1}^{-1} \frac{\partial \mathbf{R}_{\mathbf{y}_1}}{\partial \theta} & \mathbf{0} & \cdots & \mathbf{0} \\
			\mathbf{0} & \mathbf{R}_{\mathbf{y}_2}^{-1} \frac{\partial \mathbf{R}_{\mathbf{y}_2}}{\partial \theta}\mathbf{R}_{\mathbf{y}_2}^{-1} \frac{\partial \mathbf{R}_{\mathbf{y}_2}}{\partial \theta} & \cdots & \mathbf{0} \\
			\vdots & \vdots & \ddots & \vdots \\
			\mathbf{0} & \mathbf{0} & \cdots & \mathbf{R}_{\mathbf{y}_Q}^{-1} \frac{\partial \mathbf{R}_{\mathbf{y}_Q}}{\partial \theta}\mathbf{R}_{\mathbf{y}_Q}^{-1} \frac{\partial \mathbf{R}_{\mathbf{y}_Q}}{\partial \theta}
		\end{array}
		\right]
		\right\} = \sum_{q=1}^{Q}\mathbf{FIM}_{q}
	\end{align}	
	\hrulefill
	\vspace*{4pt}
\end{figure*}
\begin{align}
	CRLB &= \frac{1}{N}\mathbf{FIM}_{\mathbf{y}}^{-1} \nonumber\\
	& = \frac{1}{N}\left(\sum_{q=1}^{Q}\mathbf{FIM}_{q}\right)^{-1}.\nonumber\\
\end{align}
%
Therefore, the closed-form expression of the FIM for the $\mathrm{H}^2$AD structure is derived.

\ifCLASSOPTIONcaptionsoff
  \newpage
\fi

\bibliographystyle{IEEEtran}
\bibliography{reference}
\end{document}